\documentclass[times,10pt,twocolumn]{article} 
\usepackage{latex8}
\usepackage{times}

\usepackage{url}
\usepackage{graphicx}
\usepackage{moreverb}

\newcommand{\etal}{et al.\@}
\newcommand{\myFig}[1]{Figure~\ref{#1}}

\newcommand{\refSec}[1]{sec.~#1}
\newcommand{\refPage}[1]{p.~#1}

\newcommand{\myQtdIn}{qtd.\@{} in}

\newcommand{\citeQuotedFromTwo}[2]{\myQtdIn{} \cite[#1]{#2}}

\newcommand{\myquote}[2]{}
\newenvironment{tinyquote}{\begin{myquoteenv}\footnotesize}{\end{myquoteenv}}
\newenvironment{myquoteenv}{
  \begin{list}
    {}
    {\leftmargin=1em\rightmargin=\leftmargin}
    \item[]
    }{\end{list}}

\newenvironment{mydescription}{
  \begin{list}
    {}
    {\leftmargin=0em\itemindent=2em}
  }{\end{list}}

\newcommand{\mySection}[1]{\vspace{-1mm}\Section{\vspace{-2mm}#1}}
\newcommand{\mySubSection}[1]{\vspace{-1mm}\SubSection{\vspace{-2mm}#1}}

\begin{document}

\title{
  Towards a Process for Developing Maintenance Tools in Academia
}
\author{
  Holger M. Kienle and Hausi A. M\"uller\\
  University of Victoria\\
  Victoria, Canada\\
  {\tt \{kienle,hausi\}@cs.uvic.ca}
}
\date{}
\maketitle
\thispagestyle{empty}

\begin{abstract}
  Building of tools---from simple prototypes to industrial-strength
  applications---is a pervasive activity in academic research. When
  proposing a new technique for software maintenance, effective tool
  support is typically required to demonstrate the feasibility and
  effectiveness of the approach. However, even though tool building is
  both pervasive and requiring significant time and effort, it is
  still pursued in an ad hoc manner. In fact, little research has
  addressed the question how to make tool building in academia more
  disciplined, predictable and efficient.  In this paper, we address
  these issues by proposing a dedicated development process for tool
  building that takes the unique characteristics of an academic
  research environment into account. We first identify process
  requirements based on a review of the literature and our extensive
  tool building experience in the domain of maintenance tools. We then
  outline a process framework based on work products that accommodates
  the requirements while providing needed flexibility for tailoring
  the process to account for specific tool building approaches and
  project constraints. The work products are concrete milestones of
  the process, tracking progress, rationalizing (design) decisions,
  and documenting the current state of the tool building project.
  Thus, the work products provide important input for strategic
  project decisions and rapid initiation of new team members.
  Leveraging a dedicated tool building process promises tools that are
  designed, build, and maintained in a more disciplined, predictable
  and efficient manner.
\end{abstract}

\mySection{Introduction}\label{sec:Intro}

Typical research in the domains of software maintenance and reverse
engineering proposes techniques to improve the comprehension of
software systems. In order to evaluate a proposed technique and to
demonstrate its feasibility many researchers implement a
tool.
Such a tool can range from a simple proof-of-concept prototype to a
fully-fledged, industrial-strength application.

This paper explores the current state of tool building practices in
academia with the aim to improve upon the state-of-the-art. This topic
is worthwhile to address because even though tool building is a
popular technique to validate research in software engineering, it is
neither simple nor cheap to accomplish. For example, tools such as the
Rigi and Moose reverse engineering environments require significant
resources to develop and maintain. Rigi is now being maintained for
over a decade and this effort has been accomplished by a number of
students at the Master and Ph.D. level as well as occasionally by
dedicated staff members. In this context the resulting constant
turn-over of (unexperienced) developers is a concern. Nierstrasz
\etal{}, who have developed the Moose reengineering tool, observe that
``crafting a tool requires engineering expertise and effort, which
consumes valuable research resources'' \cite{NDG:FSE:05}. The transfer
of engineering expertise in the domain of tool building has to be
addressed within each development project. Furthermore, it also should
be addressed within the tool building community in order to
communicate the state-of-the-art and to further improve upon
it. Currently, the communication of engineering expertise to the
community is not sufficiently rewarded: research contributions in
conferences and journals are measured by novelty, not by synthesis of
existing work and experiences.

Having a process when building a maintenance tool is especially
desirable if it is a long-running project that has a significant
turnover of (student) developers. For longer-term research this is
often the case. Furthermore, to show the effectiveness of these tools
they should be sufficiently stable and mature to serve in larger-scale
(industrial) user studies. Building a high-quality tool without
following a process exposes the research project to unnecessary risks.

Improving on the current tool building practice promises tools that
are designed, build, and maintained in a more disciplined, predictable
and efficient manner. Furthermore, practices can emphasize certain
non-functional tool requirements such as usability and adoptability,
or a certain approach to tool building such as component-based
development. In this paper we advocate to employ an explicit tool
building process to raise the state-of-the-art.

This paper is organized as follows. Section~\ref{sec:Background} gives
further background on tool building in academia, concluding that it is
executed in an ad hoc manner and that a suitable process can improve
upon the state-of-the-art. We explore then two more disciplined
approaches to tool building that have been pursued in the construction
and maintenance of SHriMP and TkSee.
Section~\ref{sec:Requirements} presents the requirements that are
desirable for a tool building process in an academic
environment. These requirements have been distilled with the help of a
literature survey as well as from our own tool building experiences.
The identified requirements can, on the one hand, provide important
background information and constraints for developing an appropriate
tool-building method, and, on the other hand, serve as evaluation
criteria to judge the efficacy of a proposed tool-building method.
Section~\ref{sec:Process} is a first step towards a dedicated process
for building tools. We advocate a flexible process framework based on
work products that can be easily tailored to accommodate different
needs. Importantly, the work products that we are proposing satisfy
our identified process requirements.
Section~\ref{sec:Conclusions} closes the paper with conclusions and
future work.

\mySection{Background and Related Work}\label{sec:Background}

\myquote{We do not promote RUP for all projects, CMM level 5 for all
  organizations, XP for all teams, or object-orientation for all
  applications. Each problem, organization, and project has its own
  characteristics, requiring a range of techniques and
  strategies---but never none!}{Laplante and Neill \cite{LN:Queue:04}}

Research in tool support for software maintenance is actively pursued
by many academics in software engineering. In the following, we view
such tools as \emph{maintenance tools} that offer functionalities for
the comprehension and analysis of a target system with the goal to
support engineers in the performance of maintenance tasks. Examples of
such tools are software visualizers, bug trackers and slicers as well as
recommender systems, search and metrics engines. In this
paper we focus mostly on reverse engineering and graph-based software
visualization tools because it is our main expertise and interest.
Given that the construction of maintenance tools is a pervasive
activity in academic research, surprisingly little work has focused on
how tools are built and how to improve upon the current practice.

Some researchers have published experiences about tool building.
For example, Lanza describes his experiences with the CodeCrawler
software visualizer \cite{Lanza:CSMR:03}, discussing CodeCrawler's
architecture (composed of three subsystems: core, metamodel, and
visualization engine), the visualization engine (realized by extending
the HotDraw framework), and desirable interactive mechanisms for
usability. Furthermore, he distills lessons learned for all of the
discussed issues. He observes that ``to our knowledge there is no
explicit work about the implementation and architecture of reverse
engineering tools, and more specifically about software visualization
tools.''
Gu\'eh\'eneuc describes his use of design patterns and Java language
idioms when constructing the Ptidej tool suite
\cite{Gueheneuc:BSUP:05}.
The special issue on Experimental Software and Toolkits (EST)
\cite{Brand:EST:07} of Elsevier's \emph{Science in Computer
  Programming\/} journal is devoted to the description of academic
research tools. One tool of the special issue, g$^4$re, falls within
the maintenance domain \cite{KMP:EST:07}.

Besides the published tool building experiences discussed above,
researchers have also identified requirements for maintenance and
reverse engineering tools \cite[\refSec{3.2}]{Kienle:UVic:06}. For
example, Tichelaar discusses six requirements for reengineering
environments \cite[\refSec{5.1}]{Tichelaar:UB:01}. Wong has distilled
23 high-level and 13 data requirements for software understanding and
reverse engineering tools \cite{Wong:UVIC:99}. He also recommends to
``summarize and distill lessons learned from reverse engineering
experience to derive requirements for the next generation of tools.''

The growing interest of researchers to report their tool building
experiences is a positive development that should be further
strengthened within the research community. However, so far the focus
is almost exclusively on documenting engineering expertise, but not on
process.
Unfortunately, most researchers do not report at all about their tool
development process. It seems that researchers often develop their
tools by themselves and are the only users of the tool during and
after development. Tools are then evaluated with a case study or
anecdotal evidence. As a result, the building of tools resembles a
{\em craft} rather than {\em professional engineering}
\cite{Shaw:IEEESW:90}.

One approach to professionalize tool building is to follow a
process. Indeed, all software development projects should use a
well-defined process---tool-building in an academic environment is no
exception to this rule. According to Kruchten, without such a process
the ``development team will develop in an ad hoc manner, with
successes relying on the heroic efforts of a few dedicated individual
contributors. This is not a sustainable condition''
\cite[\refPage{15}]{Kruchten:99}. A process should provide guidance on
what work products should be produced when, how, and by whom. A
well-defined process enables a repeatable and predictable way of
building software systems.
There are a few examples of researchers that touch on process aspects
when relating their experiences.
The following sections give examples of two approaches to tool
building---SHriMP and TkSee---that can provide valuable input towards
identifying process requirements. These can in turn be used when
defining a suitable process for the domain of academic tool building.
However, to our knowledge, no full process has been proposed so
far. The closest research that we are aware of is Chirouze \etal{}'s
work on Extreme Researching (XR), which is a process specifically
designed for applied research and development in a distributed
environment that has to cope with changing human resources and rapid
prototyping \cite{CCM:SPE:05}. It is an adaptation of Extreme
Programming (XP) and has been developed by Ericsson Applied Research
Laboratories to support distributed telecommunications research.
XR is based on several core principles taken from XP (short
development cycles, test-driven development, collective ownership, and
discipline) and encodes them in a set of activities (e.g., remote pair
programming, unit testing, collective knowledge, coding standards,
frequent integration, and metaphor). Dedicated tool support is
available for XR with a web-based portal. Based on three projects, the
authors of XR estimate that their process has yielded an increase of
output of around 24\% and reduced project-overrun time on average by
half.

\mySubSection{Tool Building in SHriMP and TkSee}

The development history of the \textbf{SHriMP} program comprehension
tool is one of the few more comprehensive sources that allow us to
infer requirements for tool-building.

The SHriMP project is an example of building a tool that has been
continuously refined and evaluated, following an iterative approach of
designing and implementing tools, which has been proposed by the
team's leading professor
\cite[\refSec{11.1.1}]{Storey:SFU:98}. Evolving a tool such as SHriMP
is a major research commitment, involving several students at the
Ph.D. and master level at any given time. SHriMP had several major
implementations. The first implementation was based on Rigi (with
Tcl/Tk and C/C++), followed by a Java port. The Java version of SHriMP
was then re-architected in a component-based technology, JavaBeans, to
facilitate closer collaboration between reverse engineering research
groups.

The Rigi-based version of SHriMP was evaluated and improved based on
two user studies with 12 and 30 participants, respectively
\cite{SWFHM:WCRE:96} \cite{SWM:WCRE:97}. The results of both user
studies helped to improve SHriMP by identifying several shortcoming.
Wu and Storey have also published the results of their first Java port
of SHriMP; they state that ``during the development of this prototype,
we took into consideration the knowledge from previous prototypes and
empirical evaluations'' \cite{WS:CASCON:00}. Even though they do not
provide further details, we can infer that their development process
is (1) prototype-based, (2) iterative, and (3) based on feedback from
empirical evaluations.

The \textbf{TkSee} search tool (written in Tcl/Tk and C++) is an
example of a tool that has been improved based on the feedback
obtained from developers in industry.
Early versions of TkSee were delivered to users at Mitel in 1996. At
the time, the tool was used by relatively few user, which used only a
small subset of its features. Generally, users were reluctant to adopt
new tools. Lethbridge and Herrera describe TkSee's development process
as follows \cite[\refPage{89}]{LH:CSER:01}:
\begin{tinyquote}
  ``TkSee had been developed in a university research environment
  following an informal and opportunistic development process.
  Features had been added by students and researchers when they had
  had bright ideas they wished to experiment with, hence it lacked
  complete documents describing its requirements, its design (except
  that of its database architecture \cite{LA:UoO:97}) and how to use
  it.
  There was considerable staff turnover among TkSee developers because
  many were students. \dots{} The newer staff was often not able to
  understand the tool or the purpose of certain features.
\end{tinyquote}
To improve TkSee, feedback was obtained from field observations as
well as formal user studies using think-aloud protocol and
videotaping. The results were then communicated to the tool developers
in a report and sessions with video clips to ``emphasize certain
points and convince them of the seriousness'' of the (usability)
problems.
Similar to SHriMP, we can conclude that tool development had at least
one iteration that improved the tool based on feedback from a user
study. Furthermore, the use of Tcl/Tk would facilitate rapid
prototyping of TkSee.

The approaches to tool building of both SHriMP and TkSee provide
valuable input to identify requirements that a dedicated process for
tool building should exhibit.

\mySection{Process Requirements}\label{sec:Requirements}

This section introduces requirements for a software process to develop
maintenance tools.
A dedicated development process has to accommodate the particular
characteristics and constraints of the target domain. For instance,
tool development in an academic research environment often is ad hoc
and unstructured. Tools are often constructed by students that have
only a few years of programming experience, and that typically work
alone or in small teams with informal communication flow and without
an explicit process. Furthermore, they often work not closely
supervised and are evaluated based on their finished product, not on
how they have constructed it.

Researchers in the domain have reported some experiences that allow to
distill properties that an appropriate development process should
probably possess. We also draw from informal discussions with other
researchers in the maintenance domain, and from our own experiences of
developing software visualization and reverse engineering tools.
Our own experiences include the development and maintenance of the
Rigi
and Bauhaus
tools as well as the tools constructed in the context of the
Adoption-Centric Software Engineering (ACSE) project
\cite{Kienle:UVic:06}.

The following sections discuss the requirements that we have
been able to identify for a tool building process in an academic
research setting.

\mySubSection{Feedback-Based}\label{sec:Feedback}

\myquote{As soon as the end users see how their intentions have been
  translated into a system, the requirements will change. \dots{}
  Users don't really know what they want, but they know what they do
  not want when they see it.}{Philippe Kruchten
  \cite[\refPage{54}]{Kruchten:99}}

Many ideas for improvements of software systems originate from their
users. There is evidence that software development projects are the
more successful, the more {\em user-developer links} they have
\cite{KC:CACM:95}. These links are defined as the techniques or
channels that allow users and developers to exchange information;
examples are surveys, user-interface prototypes, requirements
prototypes, interviews, bulletin boards, usability labs, and
observational studies.

As with many other software development projects, often the
researchers developing a research tool have a poor initial
understanding of its requirements.  Requirements elicitation is
difficult, for instance, because the target users are ill-defined and
the tool's functionality might be not fully understood yet.
To alleviate this problem and to bootstrap the first release, Singer
and Lethbridge propose to first involve maintenance engineers (who
will later use the tool) to understand their particular needs and
problem domain before starting the design and implementation of the
tool \cite{SLVA:CASCON:97}. This can be achieved with questionnaires,
(structured) interviews, observation sessions, and (automated) logging
of tool use.
For instance, Lethbridge and Singer have used a simple, informal
approach to elicit initial tool requirements from future users
\cite{SL:IWPC:98}:
\begin{tinyquote}
  ``For the first release, we brainstormed a group of software
  engineers for their needs, and then designed, with their continued
  involvement, a tool called SEE (Software Exploration Environment).''
\end{tinyquote}

Once a research tool has reached a first beta release, feedback should
be obtained. Lethbridge and Singer follow a process with two main
phases: (1) study a significant number of target users to thoroughly
understand the nature of their work, and then (2) develop and evaluate
tools to help the target users work better.  Importantly, the second
phase involves ``ongoing involvement with [target users]''
\cite{LS:WESS:96}.
User feedback can be obtained, for instance, with (longitudinal) work
practice studies of individuals or groups. Such studies observe and
record activities of maintainers in their normal work environment,
working on real tasks.
The designers of the Augur software exploration tool, for instance,
report that ``to gain some initial feedback on Augur's effectiveness,
we have conducted informal evaluations with developers engaged in
active development'' \cite{FD:ICSE:04}.
Hundhausen and Douglas have used the finding of an ethnographic
study with students in a course (involving techniques such as
participant observation, semi-structured interviews, and videotape
analysis) to redefine the requirements for their algorithm
visualization tool \cite{HD:SV:02}.

Wong states that ``case studies, user experiments, and surveys are all
necessary to assess the effectiveness of a reverse engineering tool''
\cite[\refPage{93}]{Wong:UVIC:99}.
Many researchers conduct informal case studies of their tools by
themselves,\footnote{For instance, in a survey of twelve visualization
  tools, two were evaluated with user studies and eleven with a case
  study \cite{SCG:SoftVis:05}.} considering themselves as typical
users \cite{TW:SCT:98}. However, it is not clear whether this approach
can generate the necessary feedback to further improve a tool. A more
effective approach is user studies.
In the best case, feedback is provided by the actual or potential
future users of the tool; however, obtaining feedback from this type
of user is often impossible. As an alternative, researchers use
other---presumably ``similar''---subjects such as computer science
students. For example, Storey conducted a study with 12 students to
assess a new visualization technique introduced with the SHriMP tool.
Observations and user comments resulting from the study generated
several improvements \cite[\refSec{5.3}]{SWFHM:WCRE:96}. Storey
explains her strategy as follows:
\begin{tinyquote}
  ``We are currently planning further user studies to evaluate the
  SHriMP and Rigi interfaces. Observations from these studies will be
  used to refine and strengthen the framework of cognitive design
  issues which will, in turn, be used to improve subsequent redesigns
  of the SHriMP interface'' \cite{SFM:JSS:99}.
\end{tinyquote}
In another user study involving students, researchers did ask 13
questions about their visualization tool, sv3D, with the goal ``to
gather feedback from first time users about some of the sv3D
features'' \cite{MCS:IWPC:05}. They conclude that the users' ``answers
provided us with valuable information that supports the implementation
of new features in sv3D.''

To summarize, researchers have tried to obtain feedback from different
types of subjects (e.g., professionals who will use the tool, students
who substitute for ``real'' users of the tool, and, last but not
least, themselves) as well as with different methods (e.g., case
studies, field studies, surveys, and formal experiments).

\mySubSection{Iterative}\label{sec:Iterative}

\myquote{You should use iterative development only on projects that
  you want to succeed.}{Martin Fowler,
  \citeQuotedFromTwo{\refPage{17}}{Larman:04} }

Iterative software development processes---as opposed to processes
that follow a waterfall or sequential strategy---are an approach to
building software in which the overall life-cycle is composed of a
sequence of iterations. Boehm's spiral model, the Rational Unified
Process (RUP), and Extreme Programming are examples of iterative
software development processes. Developing software iteratively is
among the six best practices identified by Kruchten
\cite{Kruchten:99}.

The waterfall model provides minimal opportunity for prototyping and
iterative design \cite{Grudin:IEEEC:91}, but ``is fine for small
projects that have few risks and use a well-known technology and
domain, but it cannot be stretched to fit projects that are long or
involve a high degree of novelty or risk''
\cite[\refPage{75}]{Kruchten:99}.
As illustrated by Rigi and SHriMP, tool-building projects can run for
several years. Since research tools typically explore new ideas and
strive to advance the state-of-the-art, they are risky, potentially
involving new algorithms and techniques. Thus, the waterfall approach
is not suitable for the development of research tools.

Iterative software development has several benefits compared to the
waterfall model; especially important benefits in the context of tool
building are that user feedback is encouraged, making it possible to
elicit the tool's real requirements, and that each iteration results
in an executable tool release that can be evaluated.
Generally, iterative development is most suitable to cope with
requirements changes \cite{LN:Queue:04}. As in other software
projects, requirements in tool-building can change frequently (e.g.,
because of user-feedback and modified hypothesis).

A particular example of the iterative nature of tool development is
provided by the development of a schema and corresponding fact
extractor. A schema embodies a certain domain (e.g., C++ code or
software architecture), which has to be iteratively refined. Changes
in the schema trigger changes in the fact extractor, and vice versa
\cite{MW:ateM:03}.

\begin{figure}[htb]
  \begin{center}
    \fbox{\includegraphics[width=.9\columnwidth]{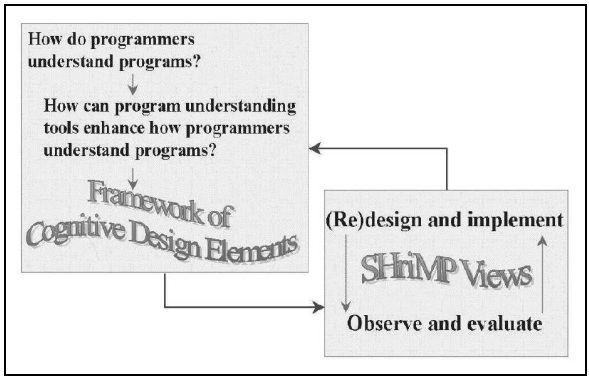}}
    \caption{Storey's tool-building process \cite{Storey:SFU:98}}
    \label{fig:StoreyProcess}
  \end{center}
\end{figure}

As an outcome of the SHriMP tool, Storey proposes an iterative
approach for designing and implementing tools \cite{Storey:SFU:98}. It
is a process consisting of ``several iterative phases of design,
development and evaluation.''  \myFig{fig:StoreyProcess} depicts a
rendering of the process iterations, illustrated with SHriMP as the
subject. There is an ``iterative cycle of design and test'' that aims
at improving a tool \cite{SFM:JSS:99}. The (initial) design of the
tool is guided by the cognitive design elements framework, which
provides a catalog of issues that should be addressed by software
exploration tools. Testing of tool design and implementation can be
accomplished by user studies. This cycle is embedded in a larger
iterative cycle for improving the framework. When adopting Storey's
process, the larger cycle could be omitted or replaced with a
different framework.

Similar to Storey, Wong argues for an iterative approach when building
reverse engineering tools. He explains the process with his proposed
Reverse Engineering Notebook:
\begin{tinyquote}
  ``The Notebook research itself undergoes continuous evolution. The
  evolution follows a spiral model \dots{} Each iteration of the
  Notebook involves several states, including studies with users,
  evaluating alternative technologies, considering compatibility
  constraints and adoption risks, validating the product, and planning
  for the next phase'' \cite[\refPage{99}]{Wong:UVIC:99}.
\end{tinyquote}

\mySubSection{Prototype-Based}\label{sec:Prototyping}

\myquote{`We prototyped our application in Visual Basic, expecting to
  rewrite it in C,' said a friend of mine recently. `But in the end we
  just shipped the prototype.'\,}{Jon Udell \cite{Udell:Byte:94}}

The construction of prototypes is common in engineering fields (e.g.,
scaled-down models in civil engineering). For software construction,
prototyping is the ``development of a preliminary version of a
software system in order to allow certain aspects of that system to be
investigated'' \cite{Schwaninger:LJ:98}.  The prototype typically
lacks functionality and does not meet all non-functional requirements.
Rapid prototyping is the use of tools or environments that support
prototype construction. Examples of technologies that facilitate rapid
prototyping are weakly-typed (scripting) languages with support for
GUI construction (e.g., Tcl/Tk and Smalltalk), tools for rapid
application development (e.g., IBM's Lotus Notes), and domain-specific
prototyping environments and languages (e.g., Matlab).

A throwaway prototype produces a cheap and rough version of the
envisioned system early on in the project---it need not be executable
and can be a paper simulation or storyboard
\cite{Cameron:CACM:02}. Such a prototype is useful to obtain and
validate user requirements and is then discarded once the initial
requirements have been established.
Exploratory prototypes are typically throwaway prototypes that are
``designed to be like a small experiment to test a key assumption
involving functionality or technology or both''
\cite[\refPage{161}]{Kruchten:99}.
As opposed to the construction of a throwaway prototype, an
evolutionary approach to prototyping continuously evolves the
prototype into the final product.

Using prototypes in software development has a number of benefits.
They are effective for defect and risk avoidance, and for uncovering
potential problems (e.g., verification of the proposed system
architecture, trying out a new technology or algorithm, identification
of performance bottlenecks, or exploration of alternative designs)
\cite{Boehm:IEEEC:99}.
Yang has used rapid prototyping to develop a tool called the
Maintainer's Assistant; he describes several benefits:
\begin{tinyquote}
  ``There are several advantages of rapid prototyping which can be
  taken to develop the Maintainer's Assistant itself. For instance,
  the system can be developed much faster by rapid prototyping, so
  that it can speed up the implementation. The user is involved in the
  process, so that he or she (mainly the builder in our system) can
  determine if the development is moving in the right direction. Rapid
  prototyping produces a model of the final system, so that the user,
  as well as builder, can see the final system in the early
  development stage'' \cite{Yang:ICSM:91}.
\end{tinyquote}

Prototyping has potential drawbacks also.
In contrast to a description of a system's behavior, executable
prototypes are well suited to represent what a system does. However,
they are less well suited to capture design rationale (i.e., why was
the system built and why in this way)---this should be documented
explicitly \cite{Schneider:ICSE:96}. There is also the threat that a
prototype that was originally planned to be thrown away is evolved
into the final product \cite{GB:IEEESW:95}. Evolutionary prototyping
is then ``merely the official name given for poor development
practices where initial attempts at development are kept rather than
thrown away and restarted'' \cite{LN:Queue:04}.

Prototypes are well suited for applied research---Chirouze \etal{} go
as far as stating that ``the idea of rapid prototyping and researching
is intrinsically linked'' \cite{CCM:SPE:05}.
Wong proposes an iterative development strategy for the building of
maintenance tools. In this approach, ``each iteration builds an
evolutionary prototype that is intended to be a solid foundation for
the next iteration'' \cite[\refPage{99}]{Wong:UVIC:99}.

Prototypes have been successfully used in the building of software
engineering tools (e.g., using Visual Basic \cite{HNHR:SCT:00}, and
Tcl/Tk \cite{SWFHM:WCRE:96}).
A tool prototype can serve as an initial proof-of-concept. Such a
prototype does not need to fulfill certain requirements. For instance,
a prototype can demonstrate the usefulness of a novel algorithm
without meeting the scalability requirements of a production tool.
However, if user studies are based on the prototype, it might have to
meet much higher quality standards.  For example, Storey \etal{}
report that
\begin{tinyquote}
  ``The first prototype of the SHriMP interface was implemented in
  Tcl/Tk. Tcl/Tk is a scripting language and user interface library
  useful for rapidly prototyping graphical interfaces. However, its
  graphics capabilities are not optimized for efficiently displaying the
  large graphs typical of software systems. The second prototype has
  been implemented using Pad++, a graphics extension for Tcl/Tk.
  Pad++ is highly optimized for efficiently displaying large numbers
  of objects and smoothly animating the motions of panning and
  zooming'' \cite{SFM:JSS:99}.
\end{tinyquote}

\mySubSection{Other Requirements}\label{sec:ProcessOtherReq}

Based on our experiences, we believe that the following requirements
should be considered for a tool-building process as well:

\begin{mydescription}
\item[\bf lightweight:] The notion of a lightweight process is not clearly
  defined. However, evidence of a lightweight process is that it
  strives to minimize (intermediate) artifacts such as vision
  statement, use-case model, and status assessment, recognizing that
  these artifacts are not the goal of a process, but a means to an
  end. Martin says, ``a good process will decrease the demand for such
  intermediate artifacts to the barest possible minimum''
  \cite{Martin:OM:01}.
  Extreme Programming has four core values from which twelve practices
  are derived; one of these values is simplicity \cite{Beck:00}. The
  SEI's Personal Software Process Body of Knowledge states that ``a
  process description should be brief and succinct'' \cite[Key Concept
  1.1.1]{PMCS:SEI:05}. A goal of configuring RUP is that the process
  ``must be made as lean as possible while still fulfilling its
  mission to rapidly produce predictably high-quality software''
  \cite[\refPage{31}]{Kruchten:99}.
  Cockburn introduces two factors that determine a method's weight:
  {\em method size} (i.e., ``number of control elements, including
  deliverables, standards, activities, milestones, quality measures,
  and so on'') and {\em method density} (i.e., ``the detail and
  consistency required in the elements'')
  \cite{Cockburn:IEEESW:00}. He says, ``a relatively small increase in
  methodology size or density adds a relatively large amount to the
  project cost'' \cite{Cockburn:IEEESW:00}.
  Agile methods are typically more lightweight than plan-driven
  approaches.

\item[\bf component-based:] Component-based software development (CBSD)
  promises lower development costs and higher productivity compared to
  implementing systems from scratch \cite{HC:01}. Researchers often
  rely on (off-the-shelf) components and products when constructing
  tools. For example, fact extraction of source code has been realized
  by leveraging Reasoning Refine, GNU GCC, Eclipse, SNIFF+ and Source
  Navigator, and visualizations of software structures have been
  implemented on top of AT\&T gaphviz, Rational Rose, Adobe
  FrameMaker, Microsoft Visio, and Web browsers.
  However, following a component-based approach for tool building has
  unique challenges. It is necessary, for instance, to evaluate
  candidate components in terms of functionality, interoperability, and
  customizability. When realizing software visualization functionality
  with components, Lanza cautions that ``reusing graph visualization
  tools and libraries like [graphviz] can break down the
  implementation time, but it can also introduce new problems like
  lack of control, interactivity and customizability. \dots{} The first
  experiments we did with external engines soon reached a limit,
  because they were not customizable and flexible enough for our
  needs'' \cite{Lanza:CSMR:03}. In order to meet such challenges, a
  process should explicitly address CBSD issues.

\item[\bf adaptive:] A process should be flexible enough to
  accommodate changing requirements of the system under
  construction. However, evolving requirements---or other changes in
  business, customer, or technological needs---might make it necessary
  to adapt the process itself during the development effort. The SEI's
  Personal Software Process Body of Knowledge states that ``a quality
  process must be easy to learn and adapt to new circumstances''
  \cite[Key Concept 5.1.4]{PMCS:SEI:05}.
  Fowler describes the adaptive nature of a process as follows, ``A
  project that begins using an adaptive process won't have the same
  process a year later. Over time, the team will find what works for
  them, and alter the process to fit'' \cite{Fowler:03}. A process can
  be adapted after an iteration as the result of a process review.

\end{mydescription}

Even though we could not find explicit evidence for these process
requirements in the literature, we believe that they should be part of
an effective tool-building approach in academia.

\mySection{Process Framework for Tool Building}\label{sec:Process}

\myquote{The most useful form of a process description will be in
  terms of work products}{Parnas and Clements \cite{PC:TSE:86}}

To advance the goal of an effective tool building approach for
academia, it is necessary to provide developers with a suitable
development process (i.e., it has to meet the process requirements of
the previous section). This process needs to encode guidelines on how
to build tools in a repeatable and predictable way.

The dilemma with proposing a process is as follows: On the one hand,
we need a development process for tool building; on the other hand,
the individual projects seem too diverse to be accommodated by a
single, one-size-fits-all process.
Each individual tool-building project has its own unique
characteristics such as the tool's requirements and functionality, the
degree of technical uncertainty, the complexity and size of the
problem, the number of developers, the background of the development
team, and so on. This is especially the case for tool-building
projects in academia, which can differ significantly from each
other.

To resolve this dilemma, we do not define and mandate a full process;
instead we are proposing a {\em process framework\/}. This framework
addresses the tool-building requirements discussed in
Section~\ref{sec:Requirements}, but needs to be instantiated by
researchers to account for the unique characteristics of their own
development projects.
The process framework is composed of a set of work products (WPs). A
WP is ``any planned, concrete result of the development process''
\cite{OOTC:97}.

The process framework's focus on WPs is inspired by IBM OOTC's
process, which is described in the book \emph{Developing
  Object-Oriented Software: An Experience-Based Approach}
\cite{OOTC:97}. It focuses on WPs because often there is agreement on
\emph{what} should be produced, but not on \emph{how} it should be
produced. The developers of the process say, ``it was decided that an
approach that standardized on work products but allowed for individual
choices on particular techniques used to produce work products was the
best fit'' \cite[\refPage{4}]{OOTC:97}. As a result, concrete WPs
provide necessary guidance for tool builders without unnecessarily
constraining them.

\mySubSection{Work Products}\label{sec:WPs}

We define seven core WPs that reflect important stages in tool
development, ranging from requirements elicitation, over architecture,
to prototype construction. The core WPs of the process framework
address specific issues of the domain. We purposely omit more generic
WPs that address issues such as implementation and testing.

Because of limited space we can only briefly describe each WP; a more
detailed description of each WP is available in the first author's
dissertation \cite[\refSec{7.2}]{Kienle:UVic:06}. We also give example
of interactions between WPs to illustrate that WPs support each other:
one WP can provide valuable input for another WP.

\begin{mydescription}
\item[\bf Intended Development Process:] This WP instantiates a suitable
  tool development process, which should meet our identified process
  requirements and account for project-specific characteristics. For
  example, a project that wants to employ components should reflect
  this practice by accommodating CBSD principles (such as guidance in
  selecting and adapting suitable components, and in assembling the
  component-based system). To support these tasks, the process
  framework already offers two WPs: Candidate Components and Tool
  Architecture.

\item[\bf Functional Requirements:] This work product captures the users'
  expectations of the tool's functionality, providing a basis for
  communication between users and developers, and enabling to estimate
  development effort. For maintenance tools, functionality will entail
  fact extraction, analysis, and visualization. These functional units
  are typically exposed in the tool's architecture (see Tool
  Architecture WP). This WP can also contain a feature list, which can
  be bootstrapped from tool comparisons published in the literature
  (e.g., \cite{BK:IWPC:01}).

\item[\bf Non-functional Requirements:] This WP describes the tool's
  quality attributes. Tools should address important domain
  requirements such as scalability, interoperability, customizability,
  usability, and adoptability. Such requirements are often difficult
  to formalize and validate. There should be a rationalization for
  each of these quality attributes on how the tool will meet them.
  Obtaining user feedback with prototypes (see Technical and UI
  Prototype WPs) can be used in such cases to clarify quality
  attributes.

\item[\bf Candidate Components:] This WP addresses the identification,
  assessment, and selection of (off-the-shelf) components and products
  that can be leveraged when building the tool. To enable a comparison
  among candidates, assessment criteria (reflecting the Functional and
  Non-functional Requirements WPs) have to be defined
  first. Documenting the rationale for selecting a certain component
  increases the confidence of the tool developers into the assessment
  and selection. Note that the selection decision may come to the
  conclusion that there is no suitable candidate component in which
  case the required functionality has to be implemented from scratch.

\item[\bf User Interface Prototype:] This WP mandates the construction of
  a UI prototype. A typical objective for a prototype is that it
  should be good enough to enable user interaction and feedback; and
  detailed and complete enough to support some kind of evaluation. The
  Functional and Non-functional Requirements WPs can be used to
  identify the tool's ``main line'' functionalities that the prototype
  should focus on. If a candidate component is available and this
  component offers scripting support then it can be used as a rapid
  prototyping environment. The development of the prototype can give
  the developers valuable first insights for the actual tool
  development.

\item[\bf Technical Prototype:] This WP is concerned with the construction
  of a prototype to explore issues related to the design,
  implementation, and architecture of the tool under
  construction. This is in contrast to the UI Prototype WP, which is
  exclusively concerned with user-interface design. Prototyping can be
  used as a risk mitigation technique to resolve or explore
  uncertainties of tool development that cannot be addressed by
  theoretical analysis or research alone. In the context of CBSD,
  exploratory prototypes are especially useful to provide input for
  the Candidate Components WP because public information about
  components can be inaccurate, misleading, or outdated.

\item[\bf Tool Architecture:] This WP documents the tool's high-level
  architecture. It is discussed in more detail in the next section.
\end{mydescription}

The full description of each WP follows a template to provide a common
structure. This template consists of description, purpose,
technique, advice and guidance, and references to related work. In the
following section, a more complete and detailed description following
the template is given.

\mySubSection{Sample Work Product: Tool Architecture}\label{sec:SampleWP}

\begin{mydescription}
\item[\bf description:] The Tool Architecture WP captures the system
  architecture of the tool under construction. It can be seen as the
  set of early, global design decisions. The architecture places
  constraints on the tool's design and implementation.
  The architecture is often visualized with diagrams, but it can also
  take the form of a textual description of non-functional
  requirements and derived architectural decisions.
  
\item[\bf purpose:] An architecture can be used to show the partitioning
  of the system into components; the communication patterns and
  mechanisms between components; the nature and services of the used
  components; etc. Without an architecture, it may be difficult to
  reason about tool properties, to communicate its design principles
  to new project members, and to maintain its (conceptual) integrity.
  The architecture can be used to reason about certain Non-functional
  Requirements of the system (e.g., performance, modifiability, and
  reusability).

\item[\bf technique:] The tool's architecture can be described as a
  conceptual architecture diagram that shows the tool's main
  components.
  Often a single customized candidate component implements the
  functionality associated with a certain tool functionality (i.e.,
  extractor, analyzer, visualizer, or repository). In the early
  project stage, a component in the diagram may indicate its intended
  functionality without revealing its realization. For example, there
  may be a component that is meant to implement repository
  functionality. Only later on in the project a decision will be made
  about the nature of the repository (e.g., local file system,
  relational database, or XML database). Still later on, the actual
  component may be chosen.
  
  \begin{figure}[htb]
    \begin{center}
      \fbox{\includegraphics[width=.95\columnwidth]{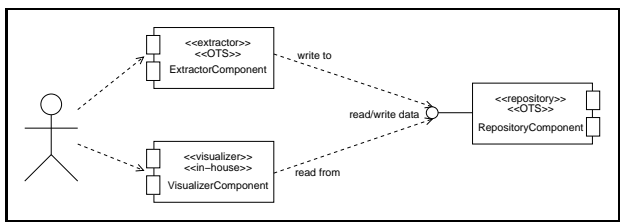}}
      \caption{A tool architecture in UML}
      \label{fig:UML}
    \end{center}
  \end{figure}
  
  Architectural diagrams are typically visualized with UML.  UML
  stereotypes can be used to convey additional meaning. For example,
  Egyed \etal{} use stereotypes to distinguish between
  \texttt{<<in-house>>} and \texttt{<<COTS>>} components
  \cite{EJB:ACSE:04}. Similarly, stereotypes can be used to identify
  the tool component functionality.
  \myFig{fig:UML} shows an example of a tool's architecture modeled in
  UML. The tool is composed of three components. The repository
  component exposes an interface (shown in UML's shorthand
  (``lollipop'') interface notation) to read and write data. The
  extractor and visualizer components use this interface to write and
  read data, respectively. Stereotypes are used to indicate the tool
  component types, and to distinguish between third-party and custom
  components.
  
\item[\bf advice and guidance:] The tool architecture can be effective to
  assess whether the resulting tool meets established design
  principles. For example, a component-based system will be more
  maintainable if it minimizes coupling among components and relies on
  open standards.
  Boehm \etal{}'s MBASE approach identifies issues---grouped into {\em
    simplifiers} and {\em complicators}---that make the development of
  component-based systems easier or harder
  \cite{Sommerville:IEEESW:05}. Simplifiers that are reflected in the
  architecture include ``clean, well-defined APIs'' and ``a single
  COTS package;'' complicators include ``multiple, incompatible COTS
  packages.''

  To come up with an architecture it is often helpful to reuse
  existing expertise in the form of reference models and
  architectures, architectural patterns, or architectural mechanisms
  \cite{Kruchten:99}. Several researchers have proposed reference
  models and architectures for reverse engineering tools (e.g.,
  \cite{LA:UoO:97} \cite{HEHHR:WCRE:95}), which can be used as a
  starting point to define a tool's architecture.

\item[\bf references:] The OOTC process defines System Architecture and
  Subsystem Model WPs; there is also a Rejected Design Alternatives
  work product that can be used to record why a certain architecture
  was not chosen.
  RUP defines two artifacts related to architecture: Software
  Architecture Document and Architectural Prototype
  \cite[\refPage{86}]{Kruchten:99}.
\end{mydescription}

\mySubSection{Discussion}\label{sec:Discussion}

The WP oriented approach of the proposed process framework has been
inspired by OOTC. However, the framework's individual WPs
are not the same as the ones described by OOTC because they are
motivated by the process constraints and characteristics of our
domain.

The process framework describes a minimum set of WPs.  A
tailored version of the framework is free to introduce additional
ones, as appropriate.
Specifically, a tool-building project that experiments with a novel or
unproven technology could introduce dedicated WPs to make sure this
risk is addressed adequately. For example, if a new tool wanted to
employ XML databases as a repository---an unfamiliar approach to the
project members---a Repository Technical Prototype WP could be
defined. This WP could prescribe tasks such as creation of a sample
schema, and benchmarking with realistic data sets.
The process framework has a number of desirable characteristics:
\begin{mydescription}
\item[\bf lightweight:] We have argued before that a tool-building
  process should be lightweight. In keeping with the requirement of a
  lightweight process, the process framework itself defines only a
  minimal set of WPs, and each WP should be as concise as possible
  (i.e., rather 2--3 pages than 10--20 pages).
  The descriptions of the WPs in the process framework are also kept
  concise (around 1000 words each) to foster a rapid deployment of the
  process.

\item[\bf adaptive:] Adaptive processes are desirable for tool-building
  projects because they often have an unstable environment (e.g., the
  composition and size of the development team can change considerably
  during its life-time).
  The WP orientation and separation of concerns of the
  process framework ``addresses the risk of losing ground when tools,
  notations, techniques, method, or process need to be adjusted by
  allowing us to vary them while maintaining the essence and value of
  completed work'' \cite[\refPage{15}]{OOTC:97}.

\item[\bf tailorable:] The framework is tailorable in the sense that
  researchers are free to extend the process framework with their own
  WPs. Researchers can also omit WPs if their
  tool-building approach has different process constraints.
  For instance, a research project may wish to incorporate an explicit
  traceability requirement for its process, or to eliminate the
  Technical Prototype WP if they have sufficient
  understanding about the technology.
  
\item[\bf separation of concerns:] The WP oriented approach leads to a
  separation of concerns: WPs are independent from tools and notation
  as well as development process and techniques.

\item[\bf reuse:] The process framework defines a reusable set of
  WPs. These can be reused by new team members to rapidly understand
  the current state of a tool project, and by other researchers to
  jump-start their own tool-building projects---thus, WPs can be seen
  as preserving valuable tool building knowledge across projects and
  people. In the best case, WPs lead to a situation where existing WPs
  are reused whenever possible, reserving the definition of new WPs
  for genuinely new contexts \cite{Cameron:CACM:02}.
\end{mydescription}

\mySection{Conclusions and Future Work}\label{sec:Conclusions}

In this paper we strive to advance academic tool building beyond the
craft stage. What is needed is a more predicable approach that
provides processes, techniques, and guidance to tool builders. The
identified process requirements (i.e., feedback-based, iterative,
prototype-based, lightweight, component-based, and adaptive) and the
outlined process framework (consisting of seven core work products)
are a first step in the direction towards the professional engineering
stage. The process framework specifically addresses the process
requirements of the domain of maintenance tool building, and an
academic environment. Importantly, we have grounded the process
requirements by analyzing the domain with a literature review of
relevant tool building experiences.

It is an open question whether the proposed process framework is
suitable for similar domains. It seems likely that the process
framework generalizes to the whole domain of software engineering tool
building and to environments that are similar to academic research
such as research labs. However, one should be careful to jump to
conclusions without a thorough investigation of the target domain. We
hope that other researchers will investigate the suitability of our
process framework in their domains.

In future work we want to apply our process framework in the
construction of maintenance tools. It seems especially promising to
first use the process on Master students implementing a relatively
small tool with well defined scope. The work products that are created
by the students during the implementation effort can be used to
document progress, record constraints, and rationalize
decision. Furthermore, the work products can be used to record
tool-building experiences as an effective means to communicate
important lesson's learned to subsequent generations of students that
begin to develop tools under similar constraints.
Lastly, we want to explore tool support for defining and evolving the
process framework and its work products. Such tool support could be
implemented on top of IBM Rational Method Composer.

\bibliographystyle{./latex8}
\bibliography{books,papers,theses,own}

\noindent \includegraphics[scale=.6]{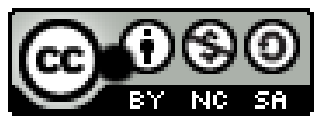}
\begin{minipage}[b]{.75\linewidth}
  {\tiny This work is licensed under a Creative Commons
    Attribution-Noncommercial-Share Alike 3.0 United States License.
    The license is available here:
    \url{http://creativecommons.org/licenses/by-nc-sa/3.0/us/}.}
  \baselineskip8pt
\end{minipage}

\end{document}